\documentclass[twocolumn,amssymb, footinbib, nobibnotes]{revtex4-1}

\usepackage{lipsum} 
\usepackage{graphicx}
\usepackage{xcolor} 
\graphicspath{{Figures/}}
\usepackage{bm}
\usepackage{braket} 
\usepackage{physics, mathtools}
\usepackage{hyperref}
\usepackage{todonotes}
\hypersetup{colorlinks,
	urlcolor=[rgb]{0,0.6,0.6},
	linkcolor=[rgb]{0,0.0,0.8}}

\renewcommand{\vec}{\mathbf}

\begin{document}

\title{Husimi function for electrons moving in magnetic fields}

\author{George Datseris}
\email{george.datseris@ds.mpg.de}
\affiliation{Max Planck Institute for Dynamics and Self-Organization, Am Fassberg 17, 37077 G\"ottingen, Germany}
\affiliation{Faculty of Physics, Georg-August-Universit\"at G\"ottingen, 37077, G\"ottingen, Germany}
\author{Ragnar Fleischmann}
\affiliation{Max Planck Institute for Dynamics and Self-Organization, Am Fassberg 17, 37077 G\"ottingen, Germany}
\affiliation{Faculty of Physics, Georg-August-Universit\"at G\"ottingen, 37077, G\"ottingen, Germany}

\date{\today}

\begin{abstract}
Husimi functions allow one to obtain sensible and useful phase space probability distributions from quantumechanical wavefunctions or classical wave fields, linking them to (semi-)classical methods and intuition. 
They have been used in several fields of physics, including electronic transport.
We show that applying Husimi functions to ballistic electron dynamics in magnetic fields needs special consideration in order for them to obey gauge invariance and energy conservation.
We therefore extend the Husimi function formalism to allow for magnetic fields making use of magnetic translation operators.
We demonstrate the application in tight-binding magneto-transport calculations in graphene nanodevices, highlighting connections with Klein tunneling.
In continuation of recent work, with this paper we further pave the way for using the Husimi function to unravel quantum transport phenomena in nanodevices. \end{abstract}

\maketitle

\section{Introduction}

The Husimi function is a useful construct that transforms a wavefunction into a (quasi-)probability distribution on the phase space.
Often denoted by $Q$, it has been introduced to quantum mechanics a long time ago by Kodi Husimi~\cite{Husimi1940}, as a smoothing of the Wigner function $W$~\cite{Wigner1932} that does not have negative probability values.
Since its introduction $Q$ has been used in various areas of physics, ranging from quantum optics~\cite{Schleich2001,Moya-Cessa2008} to ocean acoustics~\cite{Virovlyansky2012}. 
Its most prominent use however is probably in the field of quantum chaos, which tries to unravel the properties of complex quantum systems. 
For example, Husimi functions have been used to understand the structure of the eigenfunctions in paradigmatic chaotic systems like quantum maps and billiards~\cite{Nonnenmacher1998,Baecker2003,Baecker2004,Baecker2005,Toscano2008}, transport in quantum ratchets~\cite{Schanz2005}, the properties of optical microdisc lasers~\cite{Hentschel2003,Wiersig2008,Baecker2009} and even electron transport in disordered systems~\cite{Feist2006}. 

So far only very few studies, like~\cite{Feist2006, Stelzer2000}, have taken advantage of Husimi functions in the study of electron transport in nanodevices, therefore recently attempts have been made to further establish the Husimi function as a useful tool for studying nanodevices.
Mason et al.~\cite{Mason2013a, Mason2013, Mason2015} have introduced a \emph{processed Husimi map} in tight-binding models of nanodevices allowing to recover and visualize classical paths in coordinate space.
In~\cite{Datseris2019} we have used $Q$ as a probability distribution and showed how it can elucidate tunneling and scattering properties in (tight-binding) nanodevices, as well as complement the scattering matrix approach.

Usage of $Q$ for electron dynamics in magnetic fields however has been sparse so far.
In the literature we found work related to billiards (representing nanodevices)~\cite{Feist2006, Feist2009, Xu2009}, isolated atoms in magnetic fields~\cite{Jans1993, Dando1994}, solid state structures~\cite{Stelzer2000, Silva2009} and more recently quantum transport in a tight-binding nanodevice~\cite{Mason2015}.

In these references $Q$ is usually applied without any modification from its definition in the absence of magnetic fields, only Ref.~\cite{Mason2015} suggests to replace the momentum by the canonical momentum (cf.~\ref{sec:analytic}).

We surprisingly found that both approaches do not allow us to study ballistic magneto-transport because they yield Husimi functions that are typically not restricted to the energy shell and are also not gauge invariant.
In this work we show how to calculate the Husimi function for electrons moving in magnetic fields, such that these problems are eliminated.
We demonstrate this both, analytically in a continuum formulation and also numerically in a tight-binding approach similar to that of~\cite{Datseris2019}, which is suitable for simulating complex nanodevices.
We conclude by presenting an example application of our result and study a tunneling phenomenon, called Klein-Tunneling, in a graphene nanodevice in the presence of a magnetic field.

\section{Husimi functions in magnetic fields}
\subsection{Non-magnetic case}

There are several, closely related definitions of the Husimi function used in the literature.  We adopt the definition given in Ref.~\cite{Takahashi1986}, commonly used in quantum chaos, that defines the Husimi function as the magnitude of a projection of a wavefunction into the basis of minimal-uncertainty Gaussian wavepackets, or more generally as the expectation value of the density operator in the same basis.

Let $\Ket{\mathcal{W}(\mathbf{r}_0,\mathbf{k}_0, \sigma)}$ denote a Gaussian wavepacket, which in position representation and in the absence of magnetic fields reads~\cite{Takahashi1986}
\begin{equation}
\mathcal{W}(\mathbf{r}; \mathbf{r}_0,\mathbf{k}_0, \sigma) =	N_\sigma^{D/2} \exp\left( -\frac{(\mathbf{r} - \mathbf{r}_0)^2}{4\sigma^2} + i\,\mathbf{k}_0 \cdot \mathbf{r}\right)
\label{eq:wavepacket}
\end{equation}
(for $D$ spatial dimensions).
This is a Gaussian envelope in space centered at $\mathbf{r}_0$ multiplying a plane wave with wavevector $\mathbf{k}_0$. The normalization factor (in continuous space) is $N_\sigma = \left(\sigma\sqrt{2\pi}\right)^{-1}$, so that $\Braket{W|W}=1$. The key property of these wavepackets is that they minimize the uncertainty relation between position and momentum. Here $\sigma$ is the spatial uncertainty and thus is a parameter that controls the trade-off between the uncertainty in position ($\sigma$) or momentum space ($1/(2\sigma)$). 
The Husimi function $Q$  representing a system in a steady state $\psi$ is then defined as~\cite{Husimi1940, Harriman1988, Takahashi1986, Heller2018}
\begin{equation}
Q[\psi](\mathbf{r}_0,\mathbf{k}_0;\sigma)  = \left| \Braket{\psi | \mathcal{W}(\mathbf{r}_0,\mathbf{k}_0;\sigma)} \right|^2.
\label{eq:husimi_def}
\end{equation}
$Q(\vec{r}_0, \vec{k}_0)$ is a (quasi-)probability distribution on coordinates $\vec{r}_0$ and momenta $\vec{k}_0$.
In continuous space we have
\begin{equation}
\Braket{\psi | \mathcal{W}(\mathbf{r}_0,\mathbf{k}_0;\sigma)} =
\int \psi^*(\mathbf{r})  \times \mathcal{W}(\mathbf{r}, \mathbf{r}_0,\mathbf{k}_0; \sigma) \, d\mathbf{r}.
\label{eq:husimi_def_continuous} 
\end{equation}

In tight-binding systems, which are often used to represent  nanodevices, space is discritized and composed of individual lattice sites.
The projection is thus changed into a sum over the lattice sites (cf.~e.g.~\cite{Mason2013, Datseris2019})
\begin{equation}
\Braket{\psi | \mathcal{W}(\mathbf{r}_0,\mathbf{k}_0,\sigma)} = \sum_j \psi^*(\mathbf{r}_j)\times e^{-\frac{\delta \mathbf{r}_j ^2}{4\sigma^2}}e^{i\mathbf{k}_0 \cdot \mathbf{r}_j}
\label{eq:husimi_def_discrete}
\end{equation} 
where $\delta \mathbf{r}_j = \mathbf{r}_j- \mathbf{r}_0$ and $\psi(\mathbf{r}_j) \equiv \psi_j$ is the wavefunction at lattice site $j$ with position $\mathbf{r}_j$~\footnote{For each $\mathbf{r}_0$ in our simulations we use only lattice sites that are within $|\mathbf{r}_j - \mathbf{r}_0|\leq 3\sigma$, to reduce computation time.}.
Notice the complex conjugation $\psi^*$ in eqs.~\eqref{eq:husimi_def_continuous} and \eqref{eq:husimi_def_discrete}, which sometimes is omitted in the literature. While for closed systems with time-reversal symmetry it can be omitted, in the present paper it is crucial since we want to treat open systems and the magnetic field breaks the time-reversal symmetry.

\subsection{Magnetic field case}
\label{sec:analytic}
Adding a magnetic field to the description of a physical system can often be achieved by introducing \emph{minimal coupling} in the Hamiltonian, i.e. replacing the momentum $\hbar\vec{k}$ by the appropriate conjugate momentum $\hbar\vec{k}- q\vec{A}$, where $\vec{A}$ is the vector potential and $q$ the (signed) charge of the particle.
To calculate Husimi functions in a magnetic field, should $\mathcal{W}$ or eq.~\eqref{eq:husimi_def} be modified similarly?
As we mentioned in the introduction, most approaches so far did not modify $\mathcal{W}$ in any way. And Mason et al.~\cite{Mason2015} have suggested to use minimal coupling in $\mathcal{W}$ as well, i.e. $\mathbf{k}_0\to \mathbf{k}_0 - q \mathbf{A}/\hbar $.

To understand whether any (or which) modification is necessary, we will consider the energy expectation value of the Gaussian wavepackets. 
For this illustration we will use the Schr\"odinger Hamiltonian of a free charged particle in a magnetic field, which with a vector potential $\vec{A}$ becomes
\begin{equation}
H_{S,\mathbf{A}} = -\frac{\hbar ^2}{2m}\nabla ^2 + i\frac{q\hbar}{m}\mathbf{A}\cdot\nabla + \frac{q^2}{2m}\mathbf{A}^2.
\label{eq:schrodinger_magnetic}
\end{equation}
 In following computations we will set $\hbar = m = q = \sigma = 1$ for simplicity. The energy of the wavepacket of eq.~\eqref{eq:wavepacket} with $H_{S,\mathbf{A}=0}$ is 
\begin{align}
\Braket{\mathcal{W}|H_{S,0}|\mathcal{W}} =&  \int \int \mathcal{W}^* H_{S,0} \mathcal{W}\, dxdy =\nonumber \\ &
\frac{\pi}{4} \left(k_{0,x}^2+k_{0,y}^2+2\right).
\label{eq:energy_wp_noB}
\end{align}

After adding a magnetic field, we modify $\mathcal{W}$ according to Ref.~\cite{Mason2015} as
\begin{equation}
\mathcal{W}_\text{mc} := \mathcal{W}\exp(-iq\mathbf{A}\cdot\mathbf{r})
\label{eq:wavepacket_mason}
\end{equation}
(``mc'' indicating the \emph{minimal coupling}).
We again evaluate eq.~\eqref{eq:energy_wp_noB} now for non-zero vector potential and using the Landau gauge with $\mathbf{A}_L = q B (-y, 0)$ and find
\begin{align}
&\Braket{\mathcal{W}_\text{mc}|H_{S,\mathbf{A}}|\mathcal{W}_\text{mc}} = \nonumber \\ &\frac{\pi }{16} \left(5 B^2 +4 (2 B y_0+k_{0, x})^2+4 (B x_0+k_{0, y})^2+8\right).
\end{align}
Omitting the minimal coupling transformation in the Gaussian wavepacket yields a similar result. The energy expectation value is
\begin{align}
&\Braket{\mathcal{W}|H_{S,\mathbf{A}}|\mathcal{W}} = \nonumber \\ &\frac{\pi }{16} \left(B^2 +4k_{0, y}^2 + 4 (B y_0+k_{0, x})^2 + 8\right).
\end{align}

These results are surprising, as they show that the energy of the wavepackets depends on where they are centered in coordinate space, $x_0$ and $y_0$, or respectively the origin of the gauge, i.e.~the energy of the wavepackets is gauge dependent.
Using these wavepackets therefore leads to a rather unpractical and maybe even unphysical definition of the Husimi function, especially since we want to study ballistic transport where scattering is usually elastic, and the dynamics is well localized in energy around the energy shell of the Fermi energy.

We thus have to find a different approach and construct Gaussian wavepackets with energy expectation values independent of their center $\mathbf{r}_0$ or the gauge. Fortunately, translation in a magnetic field has been considered before, within the context of tight-binding theory.
It was solved in 1964 independently by Brown~\cite{Brown1964} and Zak~\cite{Zak1964}, where each one invented the \emph{magnetic translation operator group}. This operator group translates a wavefunction from one location to another one, while in the presence of magnetic fields.

We follow the approach of Brown, which states that the translation operator in magnetic fields is expressed as
\begin{align}
\hat{T}_\vec{A}(\mathbf{R}) = &\exp\left(-i\mathbf{R}\cdot (\mathbf{p} - q \mathbf{A})/\hbar \right) = \nonumber \\ & \exp\left(iq\mathbf{R}\cdot \mathbf{A}/\hbar \right) \hat{T}(\mathbf{R})
\label{eq:magnetic_translation}
\end{align}
where $\hat{T}(\mathbf{R})$ is translation operator, shifting the wavefunction from position $\mathbf{r}$ to $\mathbf{r+ R}$ in the absence of a magnetic field. Note that eq.~\eqref{eq:magnetic_translation} is formulated and only valid in the symmetric gauge, i.e. $\mathbf{A}_S = - \tfrac{1}{2} (\mathbf{r} \times \mathbf{B})$.

To construct the desired wavepacket at site $\mathbf{r}_0$ the operator $\hat{T}_\vec{A}(\mathbf{r}_0) $ must first be applied to the wavepacket of eq.~\eqref{eq:wavepacket} centererd at the origin. 
Because the expression of eq.~\eqref{eq:magnetic_translation} is valid only in the symmetric gauge, while in different scenarios different gauges are used (here we will use the Landau gauge for doing quantum transport simulations in a waveguide), the wavepacket must further undergo a gauge transform. 
This transformation is straight-forward: given a wavefunction $\psi$ in some gauge
$\mathbf{A}_S $, one changes the gauge to $\mathbf{A}_L$ by multiplying the wavefunction with  $\exp\left(i \frac{q}{\hbar}\Lambda_{S\to L}\right)$ with $\Lambda_{S\to L}$ the gauge transform. Assuming $\mathbf{B}= B\hat{z}$ the transformation from the symmetric to Landau gauge is
\begin{equation}
\Lambda_{S\to L} = \left( - \frac{B}{2} xy \right), \quad \mathbf{A}_L = \mathbf{A}_S + \nabla \Lambda_{S\to L}
\end{equation}
for $\mathbf{A}_L = -By\hat{x}$.

Sequentially applying all the transformations we arrive at what we will call \emph{magnetic Gaussian wavepacket} (in the Landau gauge and for 2 dimensional space)
\begin{widetext}
\begin{align}
\mathcal{M}_L(\mathbf{r}_0,\mathbf{k}_0, B;\sigma) & = \exp\left(i \frac{q}{\hbar}\Lambda_{S\to L}\right)\exp\left(i\frac{q}{\hbar}\mathbf{r}_0\cdot \mathbf{A}_S\right) \hat{T}(\mathbf{r}_0) \mathcal{W}(0,\mathbf{k}_0;\sigma) \nonumber \\
& = N_\sigma\exp\left( -\frac{(\mathbf{r} - \mathbf{r}_0)^2}{4\sigma^2} + i\mathbf{k}_0 \cdot (\mathbf{r}-\mathbf{r}_0)\right) \exp \left(-i \frac{q}{\hbar}\frac{B}{2}(xy -x_0y +x y_0)\right).
\label{eq:magnetic_wavepacket}
\end{align}
\end{widetext}
It is worth noting that the phase factors of eq.~\eqref{eq:magnetic_wavepacket} are not trivially related to the gauge.

Let us now reevaluate the energy expectation values (again setting $\hbar = m = \sigma = q = 1$) with this new wavepackets. We find
\begin{equation}
\Braket{H_{S, \mathbf{A}_L}|\mathcal{M}_L|H_{S,\mathbf{A}_L}} = \frac{\pi }{32}  \left(B^2 +8 \left(k_{0, x}^2+k_{0, y}^2+2\right)\right)
\label{eq:correct}
\end{equation}
which has no dependence on $x_0, y_0$ as intended. Notice that the energy does depend on the magnetic field, which is expected. It stems from the last term of eq.~\eqref{eq:schrodinger_magnetic}, which is known as the diamagnetic term. Usually this contribution is small, as it is multiplied by a factor of $q^2/m$.

\subsection{Magnetic wavepacket energy in a tight-binding system}
\label{sec:tightbind}
\begin{figure}[t!]
	\centering
		\begin{minipage}[l]{\columnwidth}
		\includegraphics[width=1\textwidth]{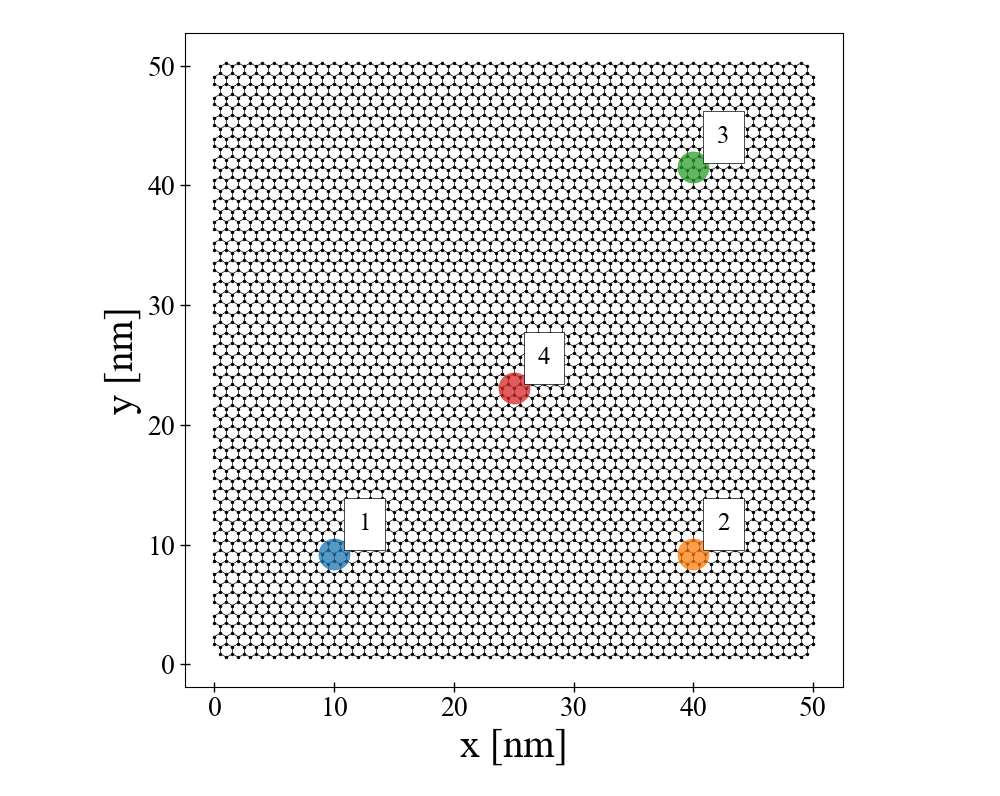}
		
	\end{minipage}
	\begin{minipage}[l]{0.8\columnwidth}
		\centering
		\resizebox{\textwidth}{!}{
			\begin{tabular}{c|ccc} 
			\multicolumn{4}{c}{Energy difference between wavepackets} \\
			\hline  
			Positions & standard & mc & magnetic \\ \hline
			(2,1) & 3.554e-15 & 4.715e-01 & 1.776e-15 \\  
			(3,1) & 1.011e-01 & 5.724e-01 & 1.776e-15 \\  
			(4,1) & 5.158e-02 & 3.181e-01 & 2.664e-15 \\  \hline
		\end{tabular}
		}
	\end{minipage}

	\caption{Energy difference between wavepackets centered at different positions, the colored region has radius $3\sigma$. For the locations 1, 2 and 3 the wavepacket centers have exactly the same distances from the edges of the sample. ``Standard'' refers to eq.~\eqref{eq:wavepacket}, ``mc'' to eq.~\eqref{eq:wavepacket_mason} and ``magnetic'' to eq.~\eqref{eq:magnetic_wavepacket}.}
	\label{fig:energy_comparison}
\end{figure}

In the continuous case we were able to evaluate the energy expectation values analytically. In the tight-binding systems couldn't, but we will numerically confirm that our approach is valid here as well, as we want to use the Husimi function in tight-binding simulations of electronic nanodevices. 
To test the validity of the magnetic wavepacket in discretized space (i.e. crystalline lattice), we create a trivial graphene rectangle, as shown in Fig.~\ref{fig:energy_comparison}, without any scalar potential.
We will compare the energy of the wavepacket when its center is located at different positions in the lattice. (We will choose $\sigma$ small enough so that we can neglect border effects.)

We set up a rectangle made out of graphene, using the software Kwant~\cite{Groth2014}.
We apply a magnetic field using the Peierls substitution (with the Landau gauge), which changes the hopping elements $t_{mn}\to t_{mn}\exp\left(  -i\frac{q}{\hbar}B \frac{y_n+y_m}{2}(x_m-x_n)\right)$, see~\cite{DatserisThesis} for details.
Then different wavepackets $\mathcal{W}(\vec{r}_0=\vec{r}_i)$ are created in the lattice centered at position $\vec{r}_i$ as indicated in Fig.~\ref{fig:energy_comparison}. We compare the energy expectation values of wavepackets defined according to eq.~\eqref{eq:wavepacket}, eq.~\eqref{eq:wavepacket_mason}, and  \eqref{eq:magnetic_wavepacket}.  We use the Hamiltonian matrix $H$ as obtained from Kwant~\cite{Groth2014}, it is a sparse matrix whose entries give either the hopping amplitudes between sites or the on-site energies. The vector product $\psi^*\cdot H\psi$ gives the energy expectation value of the wavepacket. The table of Fig.~\ref{fig:energy_comparison} compares the differences
$
|\mathcal{W}^*(\mathbf{r}_j)\cdot H\mathcal{W}(\mathbf{r}_j) - \mathcal{W}^*(\mathbf{r}_i)\cdot H\mathcal{W}(\mathbf{r}_i)|
$
for $i = 1$ and $j=2,3,4$ for the different definitions of the wavepackets.

We first see that, as expected, the energy of the unmodified Gaussian wavepacket is not independent of its position and the same holds true for the wavepacket modified by minimal coupling. Finally, we can see that the energy expectation value of the magnetic wavepacket defined above is independent of its position (within computer numerical precision).

\section{Application to a graphene nanodevice}\label{sec:application}
As an example, we will now apply the Husimi function, defined by
\begin{equation}
Q[\psi](\mathbf{r}_0,\mathbf{k}_0;\sigma)  = \left| \Braket{\psi | \mathcal{M}_L(\mathbf{r}_0,\mathbf{k}_0;\sigma)} \right|^2,
\label{eq:Mhusimi_def}
\end{equation}
to quantum magneto-transport through a graphene nanodevice, for which we chose a graphene nano-ribbon with a superimposed pn-junction as illustrated in Fig.~\ref{fig:magneticQ}.
All of our quantum transport simulations are tight binding calculations within the Landauer-B\"uttiker formalism, performed using the software Kwant~\cite{Groth2014}.
The devices are finite \emph{scattering regions} that are coupled to semi-infinite \emph{leads} (which are also graphene nano-ribbons). The \emph{modes} (eigenfunctions) of the leads enter the device and are subsequently scattered, defining the \emph{scattering wavefunctions} $\psi_m$ for each mode.
To add magnetic field we modify the hopping elements of the tight-binding system via the Peierls substitution, as described in section~\ref{sec:tightbind}.

Details on how we calculate the Husimi function in the graphene tight-binding system and the way we reduce its dimensionality can be found in Ref.~\cite{Datseris2019}. Here we will only provide a brief summary:
The Husimi function depends on 4 coordinates and is therefore hard to visualize and analyse. We will therefore examine the Husimi function only at certain vertical cuts through the ribbon. 
For the position vector $\vec{r}_0$, we thus use vertical slices at appropriate positions $x=c$, i.e. $\vec{r}_0 = \{ (c, y) : y \in [0, W]\}$. To further reduce the dimensionality, we want to exploit energy conservation and choose k-vectors on the energy shell at the Fermi-energy $E$ and parameterize the vectors on the contour by the propagation angle $\phi$.
For the wavevector $\vec{k}_0$, we use the two dimensional dispersion of graphene.
In the numerics we populate the 2D energy contour at the incoming energy $E$ with wavevectors at equally spaced angles $\phi \in [-\pi, \pi)$ and sample $Q$ using those wavevectors (graphene has six valleys of which two are not equivalent; we sample all six of them and average the result~\cite{Datseris2019}).
We have thus reduced $Q$ from a four dimensional  to a two dimensional distribution $Q(\phi, y)$.

\begin{figure}
\includegraphics[width=\columnwidth]{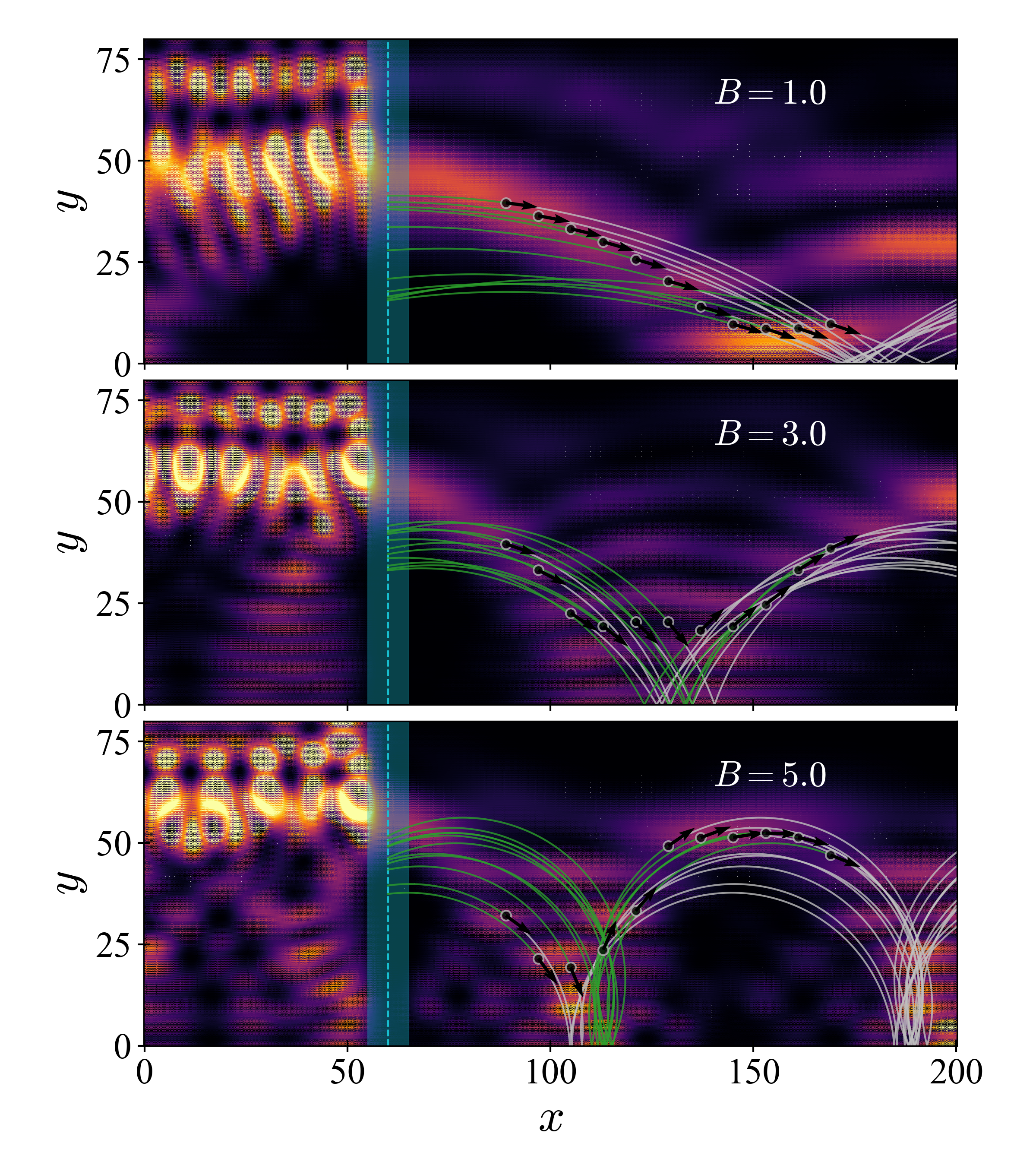}
\caption{Tunneling through a pn-junction of width $w$ in a graphene nanodevice and in the presence of magnetic fields. 
We used $L, W, \sigma, w = 200, 80, 8, 10$, $E=0.2$ eV, and $B$ as shown in the plots (in Tesla).
The scattering wavefunctions are plotted via the color map (yellow = highest amplitude), using always the mode $m=2$.
Using the Husimi function we initialize particles in a billiard (white arrows) and propagate them forwards (white color) and backwards (green color) until they reach the pn-junction (see text for more details). 
}
\label{fig:backpropagation}
\end{figure}

Strictly speaking, this step is incorrect in the presence of a magnetic field, since (as we also pointed out at eq.~\eqref{eq:correct}) the energy of the wavepackets increases with the magnetic field, and we are thus sampling an energy contour slightly different than the one with energy $E$. 
Moreover, one could argue against the usage of the 2D dispersion of graphene, since in a magnetic field it will be replaced by Landau levels.
However, we find that for the magnetic fields (up to 5 T) and incoming energies (at least 0.2 eV) that we used in this paper this process is still sufficiently accurate and yields sensible results (see Fig.~\ref{fig:magneticQ} and \ref{fig:backpropagation}).

\subsection{Husimi function of a lead}
We first present Husimi functions of a ``lead'', i.e. a pure graphene nano-ribbon that has no potential variations and is (discretely) translationally invariant in x-direction.
We are in the regime of small magnetic fields, and thus we expect semiclassically that the wavefunctions in some form follow cyclotron orbits.
The modes are ordered in by decreasing momentum in x-direction. One might therefore expect that the lower $m$, the closer the (expected) skipping orbit should be to the device's upper wall (since we use positive magnetic field).

\begin{figure*}
\includegraphics[width=\textwidth]{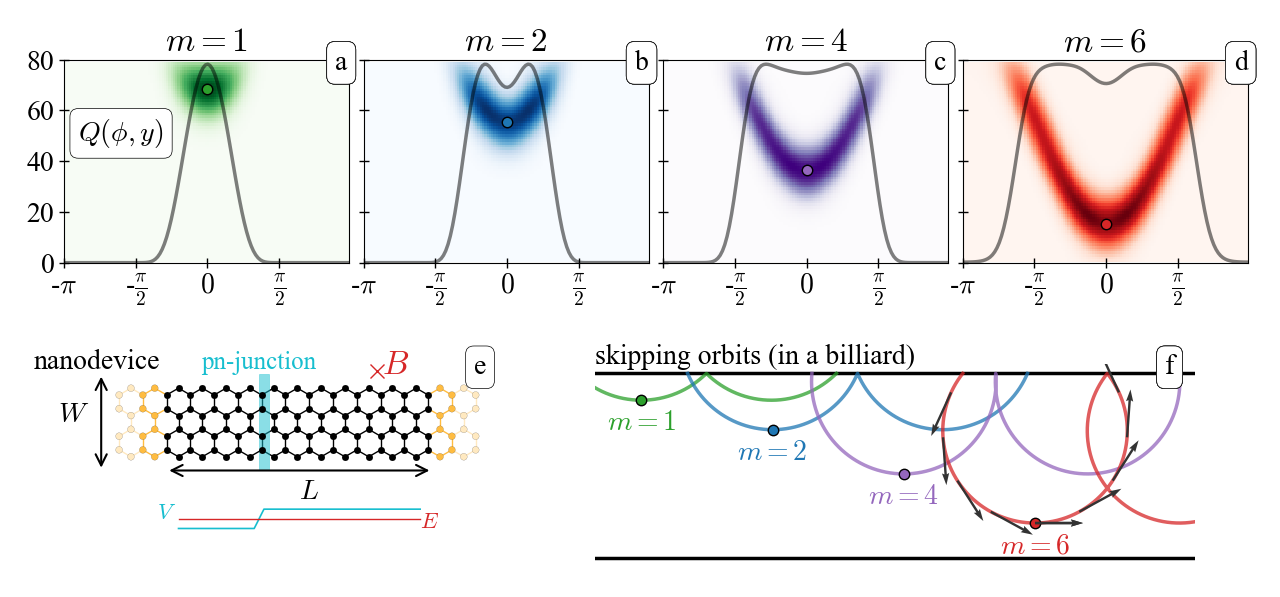}
\caption{Husimi functions in the graphene nanodevice of panel e. For producing $Q$ we used $L, W, \sigma = 200, 80, 8$ nm, $E$ = 0.2 eV and $B$ = 5 T and we did not activate the pn-junction illustrated n the sketch. The top row are the Husimi functions $Q(\phi, y)$ for the labeled incoming modes. We also overplot the marginal distribution over the angles with black color. In panel f we show characteristic skipping orbits, their color corresponding to the modes of the top row. We initialize the skipping orbits using the $y_0$ dictated by the dots in the top row. Billiard simulations done with DynamicalBilliards.jl~\cite{Datseris2017}
Below the nanodevice in panel e we show the potential landscape.}.
\label{fig:magneticQ}
\end{figure*}

The result of $Q(\phi, y)$ is shown in Fig.~\ref{fig:magneticQ}, top row, for various modes.
As we will see, the Husimi function confirms our expectations. 
Moreover it is symmetric to reflections in $\phi = 0$, since the system in closed in y-direction, and therefore positive and negative $k_y$ values have to be equally populated.
For the smallest mode $m=1$, $Q$ is localized at $\phi=0$ and $y\approx W$.
As $m$ increases, $Q$ becomes more like a ``V'' shape.
This reflects the skipping orbits moving further away from the edge.
We see that for $\phi=0$, $Q$ is localized at the lowest $y$ reached. As we increase $\phi$, $Q$ is localized at successively higher $y$ positions. 
This is exactly what a (skipping) cyclotron orbit does: as the angle shifts from zero, the direction of the particle points more towards the edge. 
We illustrate this in panel f of Fig.~\ref{fig:magneticQ}, where we simulate a magnetic billiard~\cite{Datseris2017} with the same size as the device, and use the maxima of $Q$ to initialize an electron trajectory in a magnetic field.

\subsection{pn-junction}
We now add a pn-junction in the device as illustrated in Fig.~\ref{fig:magneticQ}, i.e.~the left end of the nano-ribbon is negatively biased (the electrons populate conduction band states) and the right end of the ribbon is positively biased (the electrons populate states in the valence band). In the cyan region indicated in the figure, the bias grow linearly.
It is known that in graphene electrons inciding on such a junction show \emph{Klein tunneling}~\cite{Klein1929,Calogeracos1999,Allain2011,Cheianov2006, Datseris2019}, i.e.~at normal incidence ($\phi=0$) they are perfectly transmitted. The transmission probability decays exponentially with $\phi$, the incoming wavevector angle, as $T\approx \exp\left(-\pi k_F w \sin^2\phi\right)$ (assuming $w> 1/k_F$), which results in a collimation effect: only incoming angles that are near $\phi =0$ are successfully penetrating the junction, while larger angles are reflected.

If this $T(\phi)$ formula for Klein tunneling holds unchanged for non-zero magnetic fields appears not to be finally settled in the existing literature.
But recent publications seem to be assuming that in the semiclassical regime (i.e.~for magnetic fields significantly below the onset of the quantum Hall effect), this law for the collimation effect is still valid~\cite{Barbier2012, Patel2012, Rickhaus2015, Chen2016} and particles penetrate pn-junctions normally.

We will illustrate this effect by studying the direction the wavefunction exits the pn-junction (the direction it enters the pn-junction is just harder to visualize).
This is possible, and quite straightforward, to do with the Husimi function.
Starting with distance $3\sigma$ after the pn-junction, we measure $Q$ at vertical x-slices each with distance $\sigma$ from the previous.
We thus obtain a $Q(\phi, y)$ distribution, similar to those shown in Fig.~\ref{fig:magneticQ}, for every slice.
Then, we locate the maximum of the distribution, and obtain a value $\phi_0, y_0$ from this maximum (at every x-slice).
We use $(x, y_0, \phi_0)$ to initialize particles in a billiard with magnetic field equal to that of the device, using the software DynamicalBilliards.jl~\cite{Datseris2017}.
We then propagate these particles forwards and backwards.

It indeed seems that for magnetic fields even up to 5 Tesla, the wavefunction is exiting the pn-junction normally (we note at this point that the value of $\sigma = 8$ corresponds to an angle uncertainty of around $\Delta \phi = 0.2$, see~\cite{Datseris2019}).
This adds evidence that the transmission probability $T$ is still peaked at $\phi=0$ for non-zero magnetic fields, provided that one is within the semiclassic regime.

\bibliographystyle{ieeetr}
\bibliography{references}
\end{document}